\documentstyle[preprint,aps]{revtex}
\begin{document}
\title{Bell theorem involving all possible local measurements
}
\author{
Dagomir Kaszlikowski and Marek \.Zukowski}
\address{Instytut Fizyki Teoretycznej i Astrofizyki\\
Uniwersytet Gda\'nski, PL-80-952 Gda\'nsk, Poland}
\date{\today}
\maketitle
\begin{abstract}
The Bell theorem for a pair of two-state systems in a singlet state
is formulated for the entire range of measurement settings.
\end{abstract}

\pacs{PACS numbers: 3.65.Bz}

The Bell theorem is usually formulated with the help of the Clauser-Horne
\cite{CH} or the CHSH inequality \cite{CHSH}. These inequalities are satisfied by any
local realistic theory and are violated by quantum mechanical predictions.
They involve two apparatus settings at each of the two sides of the experiment.
However, generalisation to more than two settings at each side are possible
\cite{GAR}, \cite{BR}, \cite{ZUK}, \cite{GIS}.

There are several motivations for such generalisations. First of all new Bell inequalities 
may be more appropriate in some experimental situations, e.g., the chained Bell inequalities
can reveal violation of local realism for the Franson type experiment 
\cite{AERTS}.
Also, the academic question, why only two settings at each side, is
that always necessary, is interesting in itself. Further, many of
the currently performed quantum interferometric Bell tests did not
involve stabilisation of the interferometers at specified settings
optimal for the standard Bell inequalities, but rather involved
sample scans of the entire interferometric patterns. Thus it is useful
to have inequalities that are {\em directly} applicable to such data.

Here we present a Bell-type inequality that involves all possible settings
of the local measuring apparatus for a pair of two-state systems, which is
 always
equivalent to two spin ${1\over 2}$ particles.
The method applied is a development of the one given in \cite{ZUK}.
However, here we do not restrict ourselves to pairs of coplanar settings
(in the meaning appropriate for two Stern-Gerlach apparatuses).

Our method has two characteristic traits. The first one is that it
indeed involves the entire range of the measurement parameters. 
By this, e.g., it distinguishes itself from the limits of infinitely many
settings at each side of the so-called chained inequalities \cite{BR}, in
which not every {\em pair} of possible settings is utilised. The second one
is that the method involves the quantum prediction from the very beginning. 
As we shall see the quantum prediction determines the structure of our
Bell inequality.

In a standard Bell-type experiment one has a source emitting two particles,
each of which propagates towards one of two spatially separated
measuring devices. The particles are described by the maximally entangled
state, e.g.,
\begin{eqnarray}
&|\Psi\rangle={1\over\sqrt2}(|+\rangle_{1}|-\rangle_{2}-
|-\rangle_{1}|+\rangle_{2}),&
\label{singlet}
\end{eqnarray}
where $|+\rangle_{1}$ is the state of the first 
particle with its spin directed along the versor $\vec{z}$ of
a certain frame of reference ($-$ denotes the opposite direction), etc. 

Let as assume that every measuring device is a Stern-Gerlach
apparatus, which measures the observable $\vec{n}\cdot\vec{\sigma}$, 
where $n=a,b$ ($a$ for the first observer, $b$ for the second one),
$\vec{n}$ is a unit vector representing direction at which observer 
$n$ makes a measurement
and $\vec{\sigma}$ is a vector the components of which are standard Pauli
matrices. The family of observables $\vec{n}\cdot\vec{\sigma}$ 
covers all possible dichotomic observables for a spin ${1\over 2}$ system, endowed
with a spectrum consisting of $\pm 1$. 

In each run of the experiment every observer obtains one of the two possible
results of measurement, $\pm 1$. The probability of
obtaining by the observer $a$ the result $m=\pm 1$, when 
measuring the projection
of the spin of the incoming particle at the direction $\vec{a}$, and the result 
$m'=\pm 1$
by the observer $b$, when measuring the projection of spin of the incoming
particle at the direction $\vec{b}$ is equal to
\begin{eqnarray}
&P_{QM}(m,m';\vec{a},\vec{b})={1\over4}(1-mm'\vec{a}\cdot\vec{b}).&
\label{predictions}
\end{eqnarray} 
 
In a real experiment, however, one cannot expect that the 
observed probabilities
will follow (\ref{predictions}). Therefore, we will allow that the
interference pattern is of a reduced visibility. 
In such a case (\ref{predictions})
should be replaced by
\begin{eqnarray}
&P_{QM}(m,m';\vec{a},\vec{b})={1\over4}(1-mm'V\vec{a}\cdot\vec{b}),&
\label{predictions2}
\end{eqnarray} 
where $0\leq V \leq 1$ stands for the visibility.

From the perspective of local realism one can try to give a more
complete specification of the state of a member of the ensemble of
two-particle systems than the one given by $|\Psi\rangle$. 
The usual approach is to define a space of hidden states $\Lambda$
and a probability distribution $\rho(\lambda)$ of such states and
to represent the probability of specific results by
\begin{eqnarray}
&P_{HV}(m,m';\vec{a},\vec{b})=\int_{\Lambda}d\lambda\rho(\lambda)P_{A}(m
|\vec{a},\lambda)P_{B}(m'|\vec{b},\lambda),
\label{hiddenprob}
\end{eqnarray}
where $P_{A}(m|\vec{a},\lambda)$ 
is the probability that for given $\lambda$ and for the local observable
defined by the parameter
$\vec{a}$ the first observer obtains as a result the value
$m$ ($P_{B}(m',\vec{b},\lambda)$ plays the same role on the other
side of the experiment).

We want to check if it is possible to recover quantum mechanical
probabilities $P_{QM}$ using probabilities $P_{HV}$ based on
the assumptions of local realistic theories. We will follow
the reasoning first given in \cite{ZUK} which is 
based on the following simple geometric observation. 
Assume that one knows the components of a certain 
vector  $q$ (the {\it known} vector) belonging to some Hilbert space, 
whereas about a second
vector $h$ (the {\it test} vector) one is only
able to establish that its scalar product with $q$ satisfies
the inequality $\langle h|q \rangle < ||q||^2$. The immediate implication 
is that these two vectors cannot be equal
 $q\neq h$. 

To apply the above simple geometric fact to our case we must define appropriate
Hilbert space. Because we deal with functions 
$P_{QM}(m,m';\theta_{a},\phi_{a},\theta_{b},\phi_{b})$ and 
$P_{HV}(m,m';\theta_{a},\phi_{a},\theta_{b},\phi_{b})$
that 
depend on discrete numbers $m,m'$
and continuous variables $\theta_{n},\phi_{n}$, where
$\vec{n}=(\sin\theta_{n}\cos\phi_{n},\sin\theta_{n}
\sin\phi_{n},\cos\theta_{n})$ it is convenient to define 
the scalar product of certain two real functions $f$ and $g$ as
\begin{eqnarray}
&&\langle f| g\rangle=\sum_{m=-1}^{1}
\sum_{m'=-1}^{1}
\int d\Omega_{a}\nonumber\\
&&\times\int d\Omega_{b}
f(m,m';\theta_{a},\phi_{a},\theta_{b},\phi_{b})
g(m,m';\theta_{a},\phi_{a},\theta_{b},\phi_{b}),
\label{prodscal}
\end{eqnarray}
where $d\Omega_{n}=\sin\theta_{n}d\theta_{n}d\phi_{n}$ 
is the rotationally invariant measure
on the sphere of radius one. Our {\it known} vector is $P_{QM}$, whereas
the {\it test} one is $P_{HV}$.

One has
\begin{eqnarray}
&&||P_{QM}||^{2}=\langle P_{QM}|P_{QM}\rangle\nonumber\\
&&=(2\pi)^{2}+V^2{{4\pi}^{2}\over3}.
\label{qnorm}
\end{eqnarray} 
To estimate the scalar product $\langle P_{QM}|P_{HV}\rangle$ one has to
use the specific structure of probabilities that are
described by local hidden variables (LHV) 
(\ref{hiddenprob}). Since $P_{HV}$ is a weighted average over the hidden parameters one can make the following estimation
\begin{eqnarray}
&&\langle P_{QM},P_{HV}\rangle\leq 
\max_{\lambda\in\Lambda}\left [\sum_{m,m'=-1}^{1}
\int d\Omega_{a}\int\Omega_{b}
P_{A}(m|\vec{a},\lambda)\right.\nonumber\\
&&\left.\times P_{B}(m'|\vec{b},\lambda)
{1\over4}(1-mm'V\vec{a}\cdot\vec{b})\right].
\label{estmean}
\end{eqnarray}
Since $\sum_{m=-1}^{1}
P_{A}(m|\vec{a},\lambda)=
\sum_{m'=-1}^{1}P_{B}(
m'|\vec{b},\lambda)=1
$, the first term of (\ref{estmean}) satisfies
\begin{eqnarray}
&&{1\over4}\sum_{m,m'=-1}\int
d\Omega_{a}
\int d\Omega_{b}P_{A}(m|\vec{a},\lambda)
P_{B}(m'|\vec{b},\lambda)\nonumber\\
&&=(2\pi)^{2}.
\label{firstexp}  
\end{eqnarray}  
We transform the other term of (\ref{estmean}) to a more convenient form
\begin{eqnarray}
&&{1\over4}\sum_{m,m'=-1}^{1}\int
d\Omega_{a}\int d\Omega_{b}mm'
P(m|\vec{a},\lambda)
P(m'|\vec{b},\lambda)V\vec{a}\cdot\vec{b}\nonumber\\
&&={1\over4}\int
d\Omega_{a}\int
d\Omega_{b}
I_{a}(\vec{a},\lambda)
I_{b}(\vec{b}
,\lambda)V\vec{a}\cdot\vec{b},
\label{secondexp}
\end{eqnarray}
where $I_{n}(\vec{n},\lambda)
=\sum_{m=-1}^{1}mP_{n}(m|\vec{n},\lambda)$, and 
one has $|I_{n}(\vec{n},\lambda)|\leq1$ ($n=a,b$). 

The scalar product of two three dimensional vectors $\vec{a}$ and
$\vec{b}$ that appears in (\ref{secondexp}) can be written 
as $\vec{a}\cdot\vec{b}=
\sum_{k=1}^{3}a_{k}(\theta_{a},\phi_{a})
b_{k}(\theta_{b},\phi_{b})$,
where
\begin{eqnarray}
&\vec{n}=(n_{1},n_{2},n_{3})&\nonumber\\
&=(\sin\theta_{n}\cos\phi_{n},\sin\theta_{n}
\sin\phi_{n},\cos\theta_{n}).&
\end{eqnarray}
Therefore
(\ref{secondexp}) reads
\begin{eqnarray}
&&{V\over4}\sum_{k=1}^{3}\int
d\Omega_{a}I_{a}(\theta_{a},\phi_{a},\lambda) 
a_{k}(\theta_{a},\phi_{a})\nonumber\\
&&\times\int d\Omega_{b}
I_{b}(\theta_{b},\phi_{b},\lambda)
b_{k}(\theta_{b},\phi_{b}).
\label{tutaj}
\end{eqnarray}
We notice here that our expression is a sum of three terms,
 each of which is a product of two integrals.
 
The functions in (\ref{tutaj}) are square integrable, i.e.
integrals $\int d\Omega_{n}|I_{n}(\theta_{n},\phi_{n},\lambda)|^2$ and
$\int d\Omega_{n}|n_{k}(\theta_{n},\phi_{n})|^2$ exist (we remind that 
$|I_{n}(\theta_{n},\phi_{n},\lambda)|\leq 1$ which guarantees the
existence of the first integral). This all allows us to use formalism of
Hilbert space of square integrable functions on the unit sphere, which
we denote as $L^{2}(S^3)$.

The functions $n_{k}(\theta_{n},\phi_{n})$ fulfil the orthogonality relation
$\int d\Omega_{n}n_{k}(\theta_{n},\phi_{n})n_{l}(\theta_{n},\phi_{n})=
{4\pi\over 3}\delta_{kl}$. 
Thus, if we normalise $n_{k}$ (i.e. we divide them
by their norm, which is $\sqrt{4\pi\over3}$) we can interpret the integral
$\alpha_{k}^{n}(\lambda)=\sqrt{3\over4\pi}\int d\Omega_{k}
I_{n}(\theta_{n},\phi_{n},\lambda)n_{k}(\theta_{n}
,\phi_{n})$ as
a k-th coefficient of the {\it projection} of 
$I_{n}(\theta_{n},\phi_{n},\lambda)$ into a three dimensional 
subspace of $L^2(S^{3})$ spanned by the (normalised) basis functions 
$\sqrt{3\over4\pi}n_{k}(\theta_{n},\phi_{n})$ ($k=1,2,3$).
For later reference we will call this space $\Sigma(3)$. 
Therefore (\ref{secondexp}) transforms into
\begin{eqnarray}
&&V{\pi\over3}\sum_{k=1}^{3}\alpha^{a}_{k}(\lambda)
\alpha^{b}_{k}(\lambda).
\label{transformed}
\end{eqnarray}
Denoting the projection of $I_{n}(\theta_{n},\phi_{n},\lambda)$ into 
$\Sigma(3)$ by 
$I_{n}^{||}(\theta_{n},\phi_{n},\lambda)$ and 
using the Schwartz inequality we arrive at
\begin{eqnarray}
{\pi\over3}\sum_{k=1}^{3}\alpha_{k}^{a}(\lambda)
\alpha_{k}^{a}(\lambda)\leq {\pi\over3}||I_{a}^{||}(\cdot,\lambda)||
||I_{b}^{||}(\cdot,\lambda)||.
\label{almost}
\end{eqnarray} 

Therefore, our last step is to calculate the maximal possible value of
the norm $||I_{n}^{||}(\cdot,\lambda)||$. Since the length (norm) of a projection of a 
vector into a certain subspace
is equal to the maximal value of its scalar product with any normalised
vector belonging to this subspace,
 the norm $||I_{n}^{||}(\cdot,\lambda)||$ is given by
\begin{eqnarray}
&||I_{n}^{||}(\cdot,\lambda)||=\max_{|\vec{c}|=1}[\sqrt{3\over4\pi}\int 
d\Omega_{n}
I_{n}(\theta_{n},\phi_{n},\lambda)\sum_{k=1}^{3}c_{k}
n_{k}(\theta_{n},
\phi_{n}),
\label{mistake}
\end{eqnarray}
where $\vec{c}=(c_{1},c_{2},c_{3})$ and $|\vec{c}|=\sum_{k=1}^{3}c_{k}^2=1$. 
 Because 
$|I_{n}(\vec{a},\lambda)|\leq 1$ one has
\begin{eqnarray}
&||I_{n}^{||}(\cdot,\lambda)||\leq \max_{|\vec{c}|=1}[\sqrt{3\over4\pi}\int 
d\Omega_{n}
|\sum_{k=1}^{3}c_{k}n_{k}(\theta_{n},
\phi_{n})|].
\label{par}
\end{eqnarray}

Every vector
$\vec{c}$ can be obtained by a certain rotation of the versor $\vec{z}$.
 Such a rotation is represented by an
orthogonal matrix $\hat{O}$ belonging to the rotation group $SO(3)$. 
Therefore, (\ref{par}) can be
rewritten as
\begin{eqnarray}
&||I_{n}^{||}(\cdot,\lambda)||\leq \max_{\hat{O}}[\sqrt{3\over4\pi}\int 
d\Omega_{n}
|\hat{O}\vec{z}\cdot\vec{n}(\theta_{n},\phi_{n})|],
\label{integral1}
\end{eqnarray}
where the maximum is taken over all possible rotation matrices $\hat{O}$. Since
$|O\hat{z}\cdot\vec{n}(\theta_{n},\phi_{n})|$ is the modulus of the scalar product of 
two ordinary three dimensional vectors, it is equal to 
$|\vec{z}\cdot\hat{O}^{-1}\vec{n}(\theta_{n},\phi_{n})|$.
An active rotation of the vector $\vec{n}$ is equivalent to a (passive) 
change of the spherical coordinates. Utilising the fact that the measure 
$d\Omega_{n}$ is rotationally invariant we see that 
\begin{eqnarray}
&&||I_{n}^{||}||\leq\int d\Omega_{n}
|\sqrt{3\over4\pi}\cos\theta_{n}|=2\pi\sqrt{3\over4\pi}.
\end{eqnarray}

Therefore (\ref{almost}) is not greater then ${1\over4}(2\pi)^2$, which
with (\ref{qnorm}) and (\ref{firstexp}) gives us the following
inequalities
\begin{eqnarray}
||P^{QM}||^{2}=(2\pi)^{2}+{V^2\over 3}(2\pi)^{2}>(2\pi)^{2}+ 
{V\over 4}(2\pi)^{2}\geq \langle P^{QM},P^{HV}\rangle.
\label{damian}
\end{eqnarray}
This inequality is violated by quantum predictions provided that
the visibility V is higher then $75\%$. Please notice that
the right hand inequality is a form of a "functional" Bell inequality.
It simply gives the upper bound for the value of a certain functional
defined on the local realistic probability functions $P_{HV}$. The left
hand inequality shows that the insertion of $P_{QM}$ into the
functional Bell inequality leads to its violation provided $V>0.75$.
The characteristic trait of our functional Bell inequality
is that its form is defined by the quantum prediction $P_{QM}$.

The threshold visibility for two particle
interference to violate the inequality (\ref{damian}) is lower than
in the case of coplanar settings \cite{ZUK}, for which the critical 
visibility is
${8\over \pi^2}$. Also, it is lower than the one given recently
 by Gisin \cite{GIS}. For
his inequalities involving arbitrary many settings the threshold 
visibility equals $V={\pi\over 4}$. The chained inequalities,
for evenly spaced settings, with the number of settings going to infinity,
have the property that the critical visibility approaches in the limit 1. 

The 
question of the threshold visibility gained recently new importance.
Two-particle interferometry has been recently extended to
interference of photons which originate from independent sources \cite{ZZH}. 
Thus far the visibility is much lower than in the standard Bell-type
tests \cite{PAN}. Therefore every percentage point chopped off 
the maximal visibility for
the two-particle fringes that may still hide of local and
 realistic model
seems to be
of importance. The method presented here can easily be adapted to cases
of finite number of local parameter settings (compare e.g. \cite{ZUK}).
Therefore it can be applied directly to experimental data 
(which involve sequences of numbers, rather than continuous functions).

{\bf Acknowledgements} MZ was supported by the University of Gdansk Grant
No BW/5400-5-0264-9. DK was supported by the KBN Grant 2 P03B 096 15.
 

\begin{thebibliography}{9}
\bibitem{CH} J.F Clauser and M. A. Horne, Phys. Rev. D {\bf 10} (1974) 526.
\bibitem{CHSH} J. F. Clauser, M. A. Horne. A. Shimony and R. A. Holt, Phys. Rev. 
Lett. {\bf 23} (1969) 880.
\bibitem{GAR} A. Garuccio and F. Selleri, Found. Phys. {\bf 10} (1980) 209.
\bibitem{BR} S. L. Braunstein and C. M. Caves, in: Proc. 3rd Int.
Symp. on Foundations of quantum mechanics, eds. S. Kobayashi et. al. (Physical
Society of Japan, Tokyo, 1989).
\bibitem{ZUK} M. \.Zukowski, Phys. Lett. A, {\bf 177}, 290 (1993). 
\bibitem{GIS} N. Gisin quant-ph/9905062. 
\bibitem{AERTS} S. Aerts, P. Kwiat, J. -\AA. Larsson, and M. \.Zukowski, Phys. Rev. Lett. (to be published).

\bibitem{ZZH} M. \.Zukowski, A. Zeilinger, M. A. Horne, and A. K. Ekert,
Phys. Rev. Lett. {\bf 71} (1993) 4287
\bibitem{PAN} J. -W. Pan, D. Bouwmeester, H. Weinfurter, and A. Zeilinger,
Phys. Rev. Lett. {\bf 80} (1998) 3891
\end{thebibliography}
\end{document}